# Vie Physarale: Evaluation of Roman roads with slime mould

Emanuele Strano
Geographic Information Systems Laboratory (LaSIG). Ecole Polytechnique Fédérale de Lausanne (EPFL), Switzerland,
Urban Design Studies Unit, University of Strathclyde, Glasgow, UK

Andrew Adamatzky
Unconventional Computing Centre, University of the West of England, Bristol

Jeff Jones
Unconventional Computing Centre, University of the West of England, Bristol

**Abstract**
Roman Empire is renowned for sharp logical design and outstanding building quality of its road system. Many roads built by Romans are still use in continental Europe and UK. The Roman roads were built for military transportations with efficiency in mind, as straight as possible. Thus the roads make an ideal test-bed for developing experimental laboratory techniques for evaluating man-made transport systems using living creatures. We imitate development of road networks in Iron Age Italy using slime mould *Physarum polycephalum*. We represent ten Roman cities with oat flakes, inoculate the slime mould in Roma, wait till slime mould spans all flakes-cities with its network of protoplasmic tubes, and analyse structures of the protoplasmic networks. We found that most Roman roads, apart of those linking Placentia to Bononia and Genua to Florenzia are represented in development of *Physarum polycephalum*. Transport networks developed by Romans and by slime mould show strong affinity of planar proximity graphs, and particular minimum spanning tree. Based on laboratory experiments we reconstructed a speculative sequence of road development in Iron Age Italy.

# 1. Introduction

Developing physical, chemical and biological analogies of socio-economic processes are becoming increasingly popular nowadays because they give rise to new metaphors and uncover unique similarities. Successful examples of such cross-disciplinary fertilisation include the theory of fractal cities [7], leaf-inspired simulation of street network growth [23,6], urban theories by Alexander [5] and Salingaros [24], approaches relating urban morphology to biological morphogenesis [18], and indeed the whole branch of socio-physics [11].

Despite the overwhelming success of the bio-inspired simulation and socio-physics, the prevailing majority of publications deal with purely theoretical works and computer simulations. Almost no attempts have been made to undertake experimental laboratory comparisons between very large-scale socio-economic developments and spatio-temporal dynamics of chemical or biological systems. This could be explained by difficulties in finding a suitable experimental substrate which does not require sophisticated laboratory equipment and expensive support. A breakthrough came in 2009 when first experimental results on imitating roads networks in United Kingdom with plasmodium of slime mould *Physarum polycephalum* were published [4] followed by imitation of rail networks in Japan [30].

Plasmodium is a vegetative stage of acellular slime mould *Physarum polycephalum*. This is a single cell with many nuclei. The plasmodium fees on microscopic particles [28]. During its foraging behaviour the plasmodium spans scattered sources of nutrients with a network of protoplasmic tubes. The protoplasmic network is optimised to cover all sources of food and to provide a robust and speedy transportation of nutrients and metabolites in the plasmodium body. The plasmodium's foraging behaviour can be interpreted as computation. Data are represented by spatial configurations of attractants and repellents, and results of computation by structures of protoplasmic network formed by the plasmodium on the data sites [19,21,3]. The problems solved by plasmodium of *P. polycephalum* include shortest path [19,21], implementation of storage modification machines [3], Voronoi diagram [26], Delaunay triangulation [3], logical computing [33,2], and process algebra [25]; see overview in [3].

In the pioneering experimental works [4,30] configuration of major cities and boundaries of the countries were physically imitated by distribution of oat flakes (which attract the slime mould) and shape of agar plate (growth substrate). Reasonably good matches between protoplasmic tubes developed by *P. polycephalum* and man-made road networks were detected and experimentally verified. The findings paved a new way towards study of human transportation systems but more work is required to develop deeper conclusions, especially related to urbanization phenomena. Also, a great deal of this research was performed by computer scientists and biologists, who might lack substantial information about cities and the development of urban settlements, which may affect the interpretations of experimental laboratory results in terms of real-world transport networks. In the present paper we aim to bridge this gap and try to evaluate to what extent the emergent behaviour of *P. polycephalum* reflects structure and dynamics of human transportation systems at the macro-scale.

We focus on the well studied and historically understood *Vie Consolari* street system (VC) in the Italian peninsula. VC system is one of the first modern human transportation systems linking cities and certainly the first in Europe. VC is also a modern regional transportation network at the first stage of its development and it is also possible to trace back its evolution to observe that Rome linked all the important primordial principal cities in the Italian Peninsula. This fact is important given that VC is a transportation network that is not affected by the complex amount of changes derived by technological evolutions in transportation systems and — more generally — by the stratification of historical facts like wars, demography dynamics and natural hazards. It is derived by a top-down and self organized local organization given the complete absence of bottom-up technologies in regional planning.

Experimental simulations confirm that *P. polycephalum* matches the street network in a certain degree but we find also that there is an interesting similarities between the colonization processes in which principal streets have been built, this can suggest very relevant questions regarding self organization in the urbanization processes. It is now known that VC are now mostly overlapped by modern motorways [9] and there is a body of study dedicated in tracking the original street system in all of the Roman Empire [27]. For this study we are choosing the reconstruction proposed by [35] choosing the street systems it supposed to link the major cities in Italy at the Imperial Age (I Sec a.C.)

Vie Consolari is amongst few, if any other, ancient road networks which survived and are still in use in modern times. The Vie Consolari was the regional transportation network between major cities in Italian peninsula at the time of Roman Empire, and evolution of

Italian transport system can be traced back and hopefully replayed using configuration of "primordial" cities in the Italian peninsula.

The paper is structured as follows. In Sect. 2 we describe experimental techniques used. We analyse transport systems generated by slime mould in Sect. 3. We consider protoplasmic and man-made networks in a context of planar proximity graphs in Sect. 4. The development of approximations of Vie Consolari transport networks with a particle model of *P. polycephalum* is presented in Sect. 4. We summarise the findings of the paper in Sect. 5 and suggest further research which may reconcile the top-down requirements of efficient transport links with the bottom-up development of urbanization processes.

# 2. Experimental

We selected 11 cities $U$ for laboratory experiments (see configuration of the areas in Fig. 1):
2

1. Genua/Genova
2. Placentia/Piacenza
3. Aquileia/Venezia
4. Bononia/Bologna
5. Florenzia/Firenze
6. Ariminum/Rimini
7. Roma/Roma
8. Capua/Capua
9. Venusia/Venusia
10. Brundisium/Brindisi
11. Rheghium Reggio/Calabria

Plasmodium of *P. polycephalum* is cultivated in plastic container, on paper kitchen towels moistened with still water, and fed with oat flakes. For experiments we use 120 mm polystyrene square Petri dishes and 2% agar gel (Select agar, by Sigma Aldrich) as a substrate. Agar plates, about 2-3 mm in depth, are cut in a shape of Italy. To represent Apennine Mountains impassable for vehicular traffic we removed corresponding parts of agar plate.

The major cities $U$ are imitated by oat flakes placed in the positions of agar plate corresponding to the areas. At the beginning of each experiment an oat flake colonised by plasmodium is placed in Rome area. We undertook 28 experiments. The Petri dishes with plasmodium are kept in the dark, at temperature 22-25° C, except for observation and image recording. Periodically, we took photographs of the experimental Petri dishes with Canon D50 digital camera.

The graph $H$ of Roman road network (Fig. 1b) is constructed as follows. Let $U$ be a set of urban regions/cities; for any two regions $a$ and $b$ from $U$, the nodes $a$ and $b$ are connected by an edge $(ab)$ if there is a road starting in vicinity of $a$, passing in vicinity of $b$, and not passing in vicinity of any other urban area $c \in U$. In the case of branching – that is, a road starts in $a$, goes in the direction of $b$ and $c$, and at some point branches towards $b$ and $c$ – we then add two separate edges $(ab)$ and $(ac)$ to the graph $H$. The road graph is planar (Fig. 1b).

To generalise our experimental results we constructed a Physarum graph with weighted-edges. A Physarum graph is a tuple $P = < U, E, w >$, where $U$ is a set of urban areas, $E$ is a set edges, and $w : E \rightarrow [0,1]$ associates each edge of $E$ with a probability (or weights). For every two regions $a$ and $b$ from $U$ there is an edge connecting $a$ and $b$ if a plasmodium's protoplasmic link is recorded at least in one of $k$ experiments, and the edge $(a, b)$ has a probability calculated as a ratio of experiments where protoplasmic link $(a, b)$ occurred in the total number of experiments $k = 28$. For example, if we observed a protoplasmic tube connecting areas $a$ and $b$ in 12 experiments, the weight of edge $(a, b)$ will be $w(a, b) = \frac{12}{28}$. We do not take into account the exact configuration of the protoplasmic tubes but merely their existence. Further we will be dealing with threshold Physarum graphs $P(\theta) = < U, T(E), w, \theta >$. The threshold Physarum graph is obtained from Physarum graph by the transformation: $T(E) = \{e \in E : w(e) \geq \theta\}$. That is all edges with weights less than $\theta$ are removed.

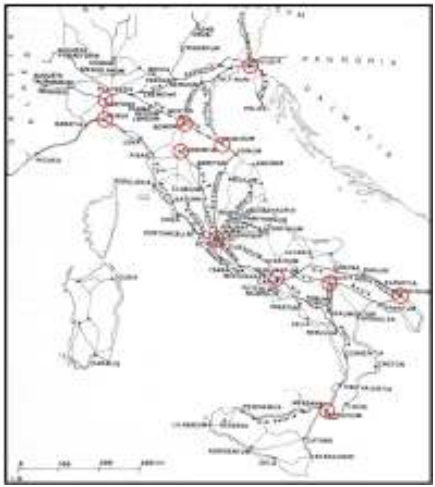 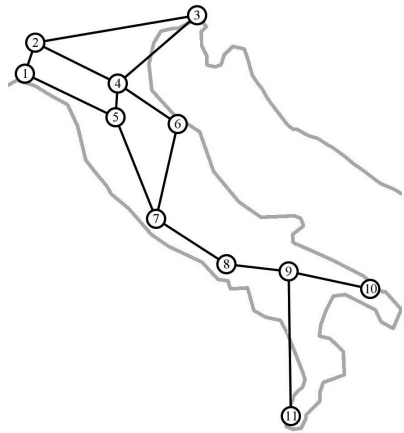

**(a)**                          **(b)** $[H]$

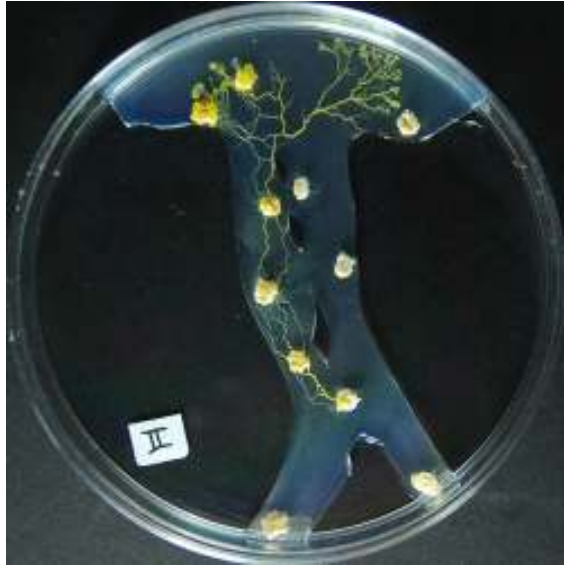

(c) [ $H$ ]

FIGURE 1. Experimental setup. (a) Outline map of Italy (from [35]) with Roman cities $U$ shown by discs. (b) motorway graph $H$ derived. (c) example of experimental setup.

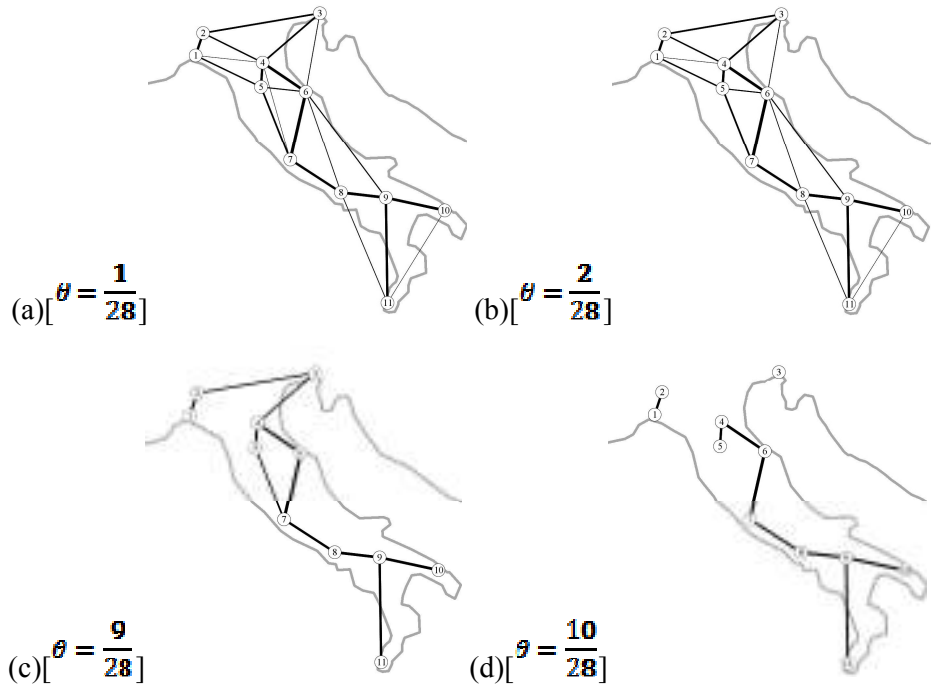

(a)[ $\theta = \dfrac{1}{28}$ ]

(b)[ $\theta = \dfrac{2}{28}$ ]

(c)[ $\theta = \dfrac{9}{28}$ ]

(d)[ $\vartheta = \dfrac{10}{28}$ ]

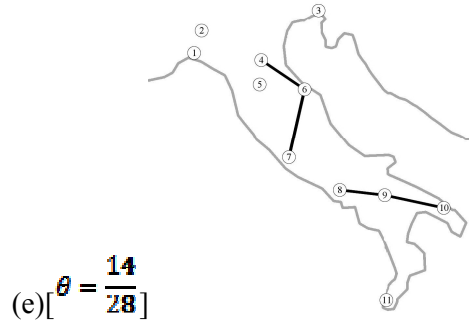

(e) $[\theta = \dfrac{14}{28}]$

**FIGURE 2. Physarum graphs $P(\theta)$ for selected values of $\theta$.**

# 3. Experimental laboratory results and their analysis

Exemplars of generalised Physarum graphs $P(\theta)$ extracted from laboratory experiments are shown in Fig. 2. We call Physarum graph $P\left(\dfrac{1}{28}\right)$ raw and Physarum graph $P\left(\dfrac{9}{28}\right)$ strong. This is because $P\left(\dfrac{1}{28}\right)$ represents protoplasmic links developed in at least one experiments. The graph $P\left(\dfrac{9}{28}\right)$ gives us more reliable representation because its edges are represented by slime mould in third of all experiments.

One raw Physarum graph $P\left(\dfrac{1}{28}\right)$ is non-planar due to edge Bononia to Roma (Fig. 2a). The graphs become planar for $\theta \geq 2$ (Fig. 2b) and remain connected till $\theta = 9$ (Fig. 2c).

Physarum graph $P\left(\dfrac{9}{28}\right)$ is the last connected graph in the series $P(\theta)$, $\theta - \dfrac{\frac{1}{28,2}}{28}, \ldots, 1$. When $\theta$ becomes 10, Physarum graphs splits into northern component: the transport link between Genua and Florenzia becomes separated from the rest of the network (Fig. 2d) and Venusia becomes isolated vertex.

Structure of Physarum graph $P(\theta)$ remains unchanged for $\theta - 10, 12$, and 13. However when $\theta = 14$ half of the cities become isolated vertexes: Genua, Aquileia, Piacentia, Filorenzia, Reghium Reggio, and only two two-node transport chains sustain (Fig. 2e). The transport chains Capua – Venusia – Brundisium and Bononia – Ariminu – Roma (segments of Via Aemelia and ViaFlamimia) are represented in half of the experiments.

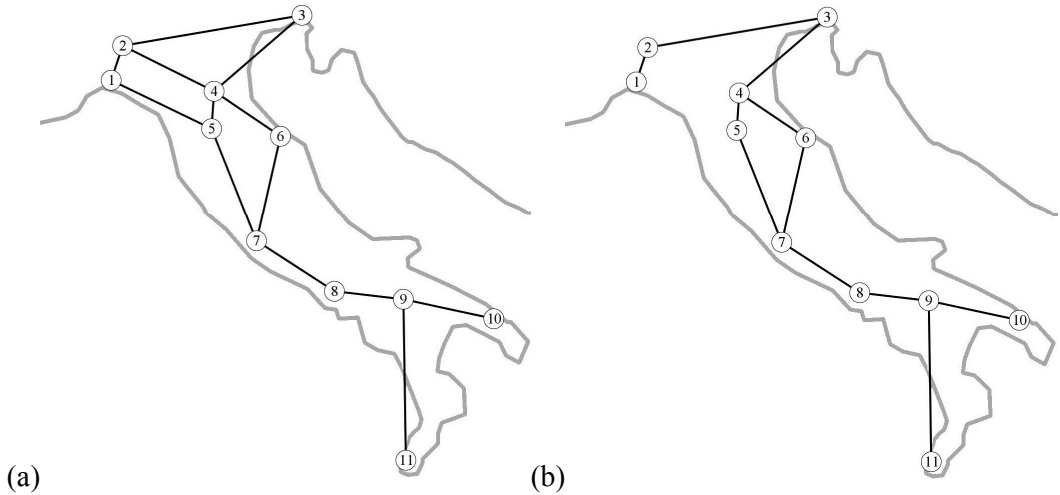

(a)                              (b)

**FIGURE 3. Intersection of road graph $H$ and Physarum graphs (a) $P\left(\frac{1}{28}\right)$ and (b) $P\left(\frac{9}{28}\right)$**

**Finding 1.** *Raw Physarum graph includes the road graph. Strong Physarum graph is included in the road graph.*

This is because $P\left(\frac{1}{28}\right) \cap H = H$ (compare Figs. 3a and 1b) and $P\left(\frac{9}{28}\right) \cap H = P\left(\frac{9}{28}\right)$ (compare Figs. 3b and 1b). The strong Physarum graph would provide perfect approximation of the road graph $H$ if roads from Piacentia to Bononia and from Genua to Florenzia were represented by protoplasmic tubes in over half of laboratory experiments.

The road graph $H$ and Physarum graphs $P(\theta)$, $\theta > \frac{2}{28}$, are planar graphs. Roman builders aimed to provide shortest transport connection between neighbouring cities while slime mould followed gradients of chemo-attractants emitted from oat flakes, representing the cities. Therefore it would be reasonable to check what is the place of the graphs $H$ and $P(\theta)$ in the family of planar proximity graphs.

A planar graph consists of nodes which are points of the Euclidean plane and edges which are straight segments connecting the points. A planar proximity graph is a planar graph where two points are connected by an edge if they are close in some sense. A pair of points is assigned a certain neighbourhood, and points of the pair are connected by an edge if their neighbourhood is empty. Here we consider the most common proximity graph as follows.

- $GG$ : Points $a$ and $b$ are connected by an edge in the Gabriel Graph $GG$ if disc with diameter $dist(a, b)$ centred in middle of the segment $ab$ is empty [10,17] (Fig. 4a).

- $RNG$ : Points $a$ and $b$ are connected by an edge in the Relative Neighbourhood Graph $RNG$ if no other point $c$ is closer to $a$ and $b$ than $dist(a, b)$ [32] (Fig. 4b).

- $MST$ : The Euclidean minimum spanning tree (MST) [22] is a connected acyclic graph which has minimum possible sum of edges' lengths (Fig. 4b).

In general, the graphs relate as $MST \subseteq RNG \subseteq GG$ [32,17,13]; this is called Toussaint hierarchy.

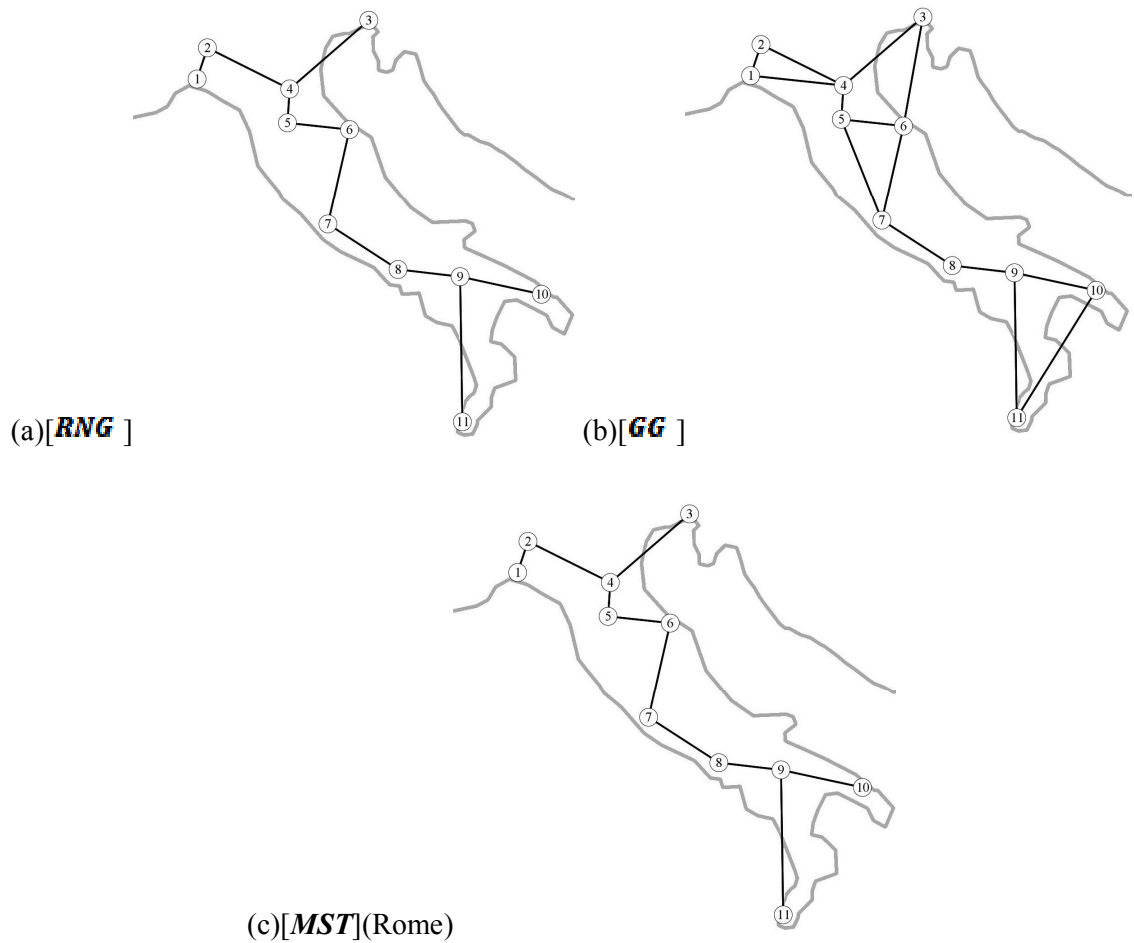

**Finding 2. $MST$ (U) = RNG (U)**

This follows directly from Figs. 4a and 4c. Equality of spanning and relative neighbourhood graph may be attributed to the fact that Italian peninsula is rather narrow yet long and therefore planar proximity graphs built on $U$ are heavily constrained.

**Finding 3. $MST \subseteq RNG \subseteq GG \subseteq P\left(\frac{1}{28}\right)$**

The raw Physarum is a sub-graph of three principle proximity graphs. This follows directly from Figs. 4 and 2a.

Strong Physarum graph $P\left(\frac{1}{28}\right)$ does not include $MST$ because this Physarum graph does not represent transport links Placentia – Bononia and Genua – Florenzia.

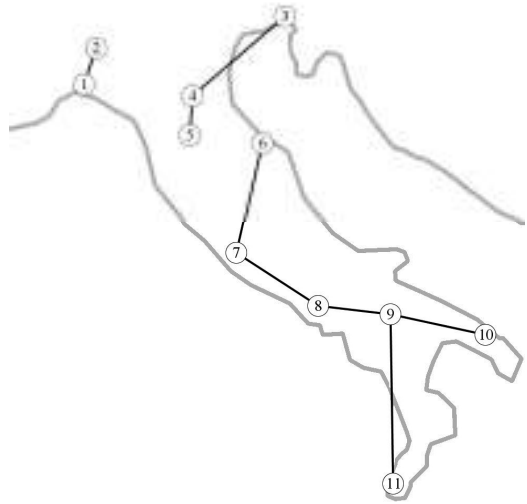

**FIGURE 5. Intersection of strong Physarum graph and minimum spanning tree.**

# 4. Simulation

The simulation model is a particle based reaction-diffusion (RD) mechanism. Unlike classical RD models, which are composed of the interactions of two simulated activator/inhibitor reactants in a diffusive environment, there is only a single representative reactant a mobile particle which senses and deposits simulated chemoattractant as it moves within a diffusive environment. The model is based on the LALI approach (Local Activation, Lateral Inhibition) to RD pattern formation [8] that was used in [14] to generate emergent dynamical transport networks.

A population of mobile particles is created and initialised on a 2D lattice configured to the experimental pattern of Italian cities. The diffusive medium is represented by a discrete two-dimensional floating point lattice. Particle positions are stored on a discrete lattice isomorphic to the diffusive lattice. Particles also store internal floating point representations of positional and orientation which is rounded to a discrete value to compute movement updates and sensory inputs. A single particle, and an aggregation of particles, is related to the *P. polycephalum* plasmodium in the following way: The plasmodium is conceptualised as an aggregate of identical components. Each particle represents a hypothetical unit of gel/sol interaction. Gel refers to the relatively stiff sponge-like matrix composed of actin-myosin fibres and sol refers to the protoplasmic solution which flows within the matrix. The *structure* of the protoplasmic network is indicated by the particle positions and the *flux* of sol within the network is represented by the movement of the particles. The resistance of the gel matrix to protoplasmic flux of sol is generated by particle-particle movement collisions.

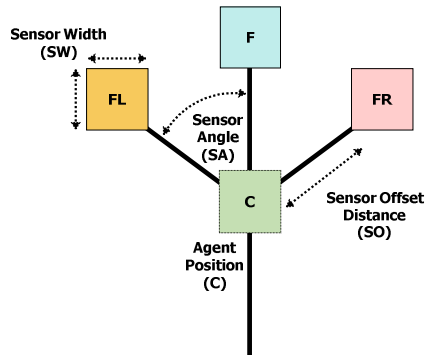

```
[Sensory stage]

- Sample chemoattractant map values
- if (F > FL) && (F > FR)
        - Continue facing same direction
- Else if (F < FL) && (F < FR)
        Rotate by RA towards larger of FL and FR
- Else if (FL < FR)
        Rotate right by RA
- Else if (FR < FL)
        Rotate left by RA
- Else
        Continue facing same direction
```

**(a)**                      **(b)**

**6. Single agent particle. (a) Morphology of agent showing sensors FL, F, FR and position C, (b) Sensory algorithm.**

The morphology of each particle is shown in Fig. 6a. The particle 'body' occupies a single cell in the lattice whose configuration is specified by a digitised greyscale image. Specific greyscale values of the image represent habitat features such as empty habitable space, boundaries and food sources. Each particle has three sensors which sample the environment at some distance away (SO — sensor offset distance, in pixels) from the particle. The offset sensor positions generate local sensory coupling of the particle population. The cohesion of the aggregate population is generated by the mutual attraction to the stimuli deposited by the particles. The coupling of particle sensors and the autocatalytic nature of the particle movement results in complex collective pattern formation and pattern evolution.

The sensory stage of the algorithm is given in Fig. 6b. Particles sense chemoattractant concentration at lattice sites covered by the three sensors. At each scheduler step particles orient towards the strongest concentration of local stimuli. Particle behaviour is very simple and is explicitly forward biased — the particle does not contemplate its current position Ð so there is an implicit emphasis on forward movement. By adjusting the SA/RA parameters a wide variety of reaction-diffusion patterning can be generated. The characteristics of the pattern formation types and its parametric evaluation were discussed in [15].

The motor stage is executed after the sensory stage. Each particle attempts to move forward in its current direction (represented by an internal state from 0-360°). If the new site is vacant, the particle occupies the new site, depositing chemoattractant (5 arbitrary units) at the new location. If the site is occupied no deposition is made and a new random orientation is selected. The particle population is iterated in a random order for both sensory and motor stages to avoid any influence from sequential updates. Each sensory and motor update, combined with environmental diffusion of chemoattractant is considered as one scheduler step.

Chemoattractant nutrient stimuli at city locations are projected to the diffusive map at every step of the scheduler (2.55 units for low nutrient concentration and 255 units for high concentration). This level of stimulus attracts the adapting virtual plasmodium and acts to constrain the evolution of the network pattern. The chemoattractant stimuli are diffused by means of a simple 3 × 3 mean filter kernel. The diffusing chemoattractant values are damped by multiplying by 0.9 in order to adjust the strength of the diffusion gradient and the distance at which the plasmodium collective can sense the nutrients.

Growth and adaptation of the particle model population is currently implemented using a simple method based upon local measures of space availability (growth) and overcrowding (adaptation by population reduction). Growth and shrinkage states are iterated separately for each particle at every two scheduler steps. For growth we assess if there are 1 to 10 particles in a $9 \times 9$ neighbourhood of a particle, and the particle has moved forwards successfully. If so, the particle attempts to divide into two if there is an empty location in the immediate neighbourhood surrounding the particle. For shrinkage we assess if there are 0 to 24 particles in a $5 \times 5$ neighbourhood of a particle. If so, the particle survives, otherwise it is annihilated.

## 4.1. Simulation Results

The topology of *P. polycephalum* transport networks is, in part, influenced by unpredictable influences on its formation, for example, the initial migration direction of the plasmodium or the presence of previously laid down protoplasmic tubes. This raises the question as to whether the topology of the networks would be more regular under idealised adaptation conditions. In order to minimise this unpredictability we initialised the simulation model with a uniform distribution of a fully grown virtual plasmodium to solely assess the effect of the spatial arrangement of nutrients (corresponding to city locations) had on the morphological adaptation of the virtual transport network. The results of an example evolution of the virtual plasmodium are shown in Fig. 7. Uniform coverage was attained by populating 50% of the habitable area of Italy with 14789 particles. The simulation was started and the collective adapted to the nutrient stimuli by shrinking in size to form a transport network connecting the city locations and avoiding the mountainous regions. Twenty runs of the simulation were performed at both low and high nutrient concentration.

We used the same method of assessing connectivity between cities as in the experimental approach and this resulted in an adjacency matrix containing the frequency connected nodes for the 20 experiments in both nutrient conditions. The connectivity graphs at different threshold weights are shown in Fig. 8 for low nutrient concentrations and Fig. 9 for high nutrient concentrations.

**Finding 4.** $MST \subset V\left(\frac{12}{20}\right)$ *at low nutrient concentration and* $MST \subset V\left(\frac{20}{20}\right) \cup$ *{Placentia, Bolonia} at high nutrient concentration.*

The higher the attractive value of nodes the more likely virtual slime mould connect the nodes with a minimum spanning tree.

**Finding 5.** $P\left(\frac{10}{20}\right) \subset V\left(\frac{20}{20}\right)$

The results demonstrate that the idealised virtual plasmodium redundantly represent generalised Physarum. Moreover, strong components Bononia — Ariminum — Roma and Capua — Venusia — Brundisium are represented by $P\left(\frac{14}{20}\right)$ and $V\left(\frac{20}{20}\right)$.

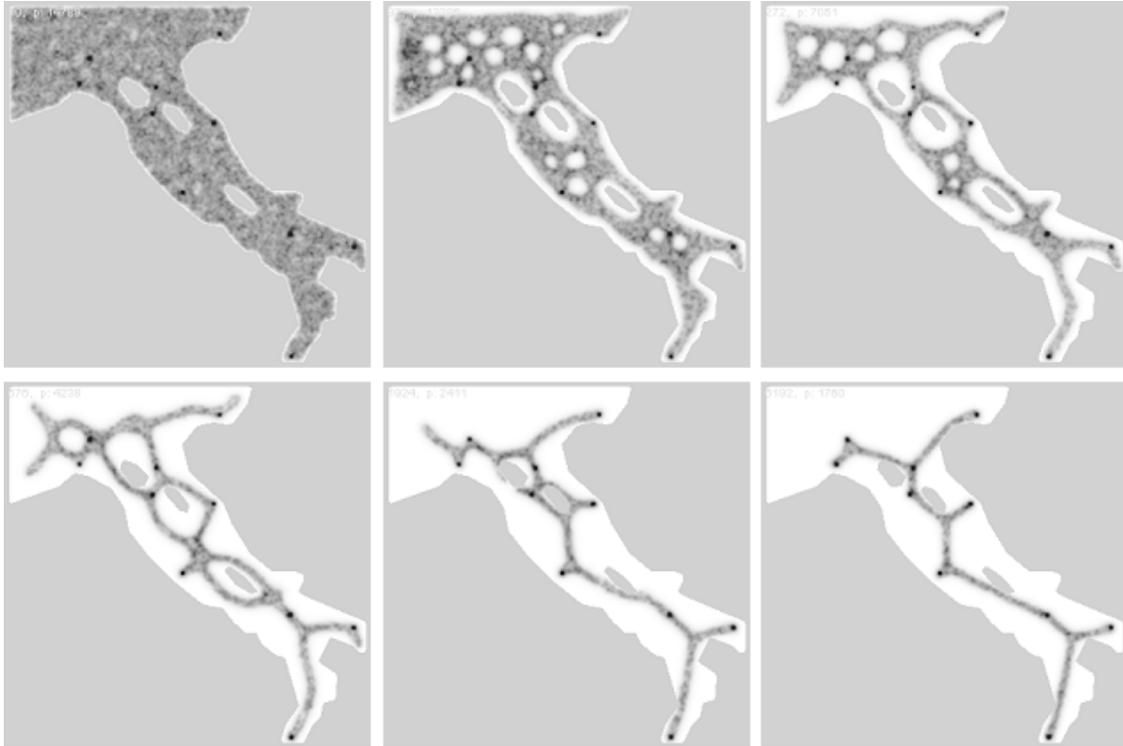

**FIGURE 7. Example of evolution of virtual plasmodium adapting its morphological pattern to form transport network connecting cities at low nutrient strengths. Image snapshots taken at 10, 91, 272, 576, 1924 and 6192 scheduler steps.**

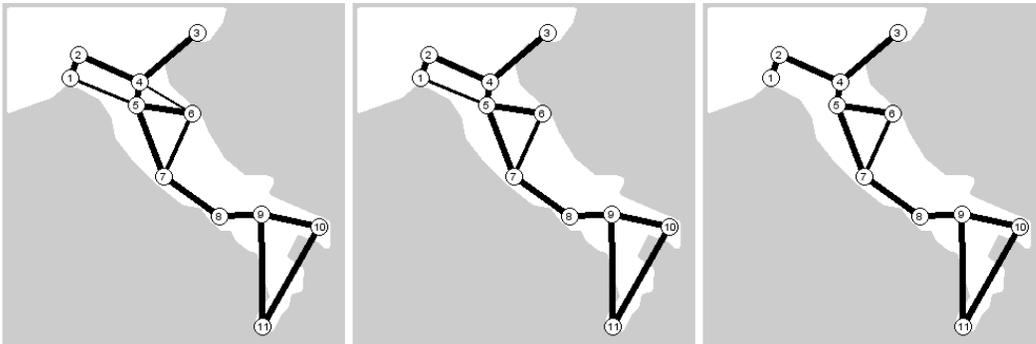

(a) $\theta = \dfrac{8}{20}$      (b) $\theta = \dfrac{10}{20}$      (c) $\theta = \dfrac{12}{20}$

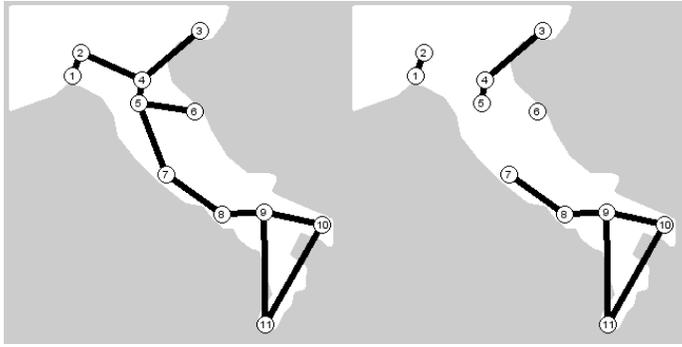

(d) $\theta = \dfrac{19}{20}$     (e) $\theta = \dfrac{20}{20}$

**FIGURE 8. Virtual plasmodium graphs $V = (\theta)$ for selected values of $\theta$ at low nutrient concentration.**

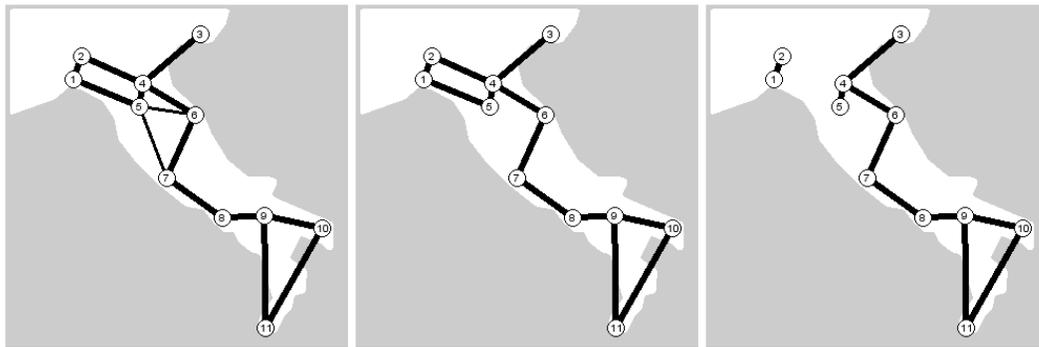

(a) $\theta = \dfrac{10}{20}$     (b) $\theta = \dfrac{19}{20}$     (c) $\theta = \dfrac{20}{20}$

**FIGURE 9. Virtual plasmodium graphs $V = (\theta)$ for selected values of $(\theta)$ at high nutrient concentration.**

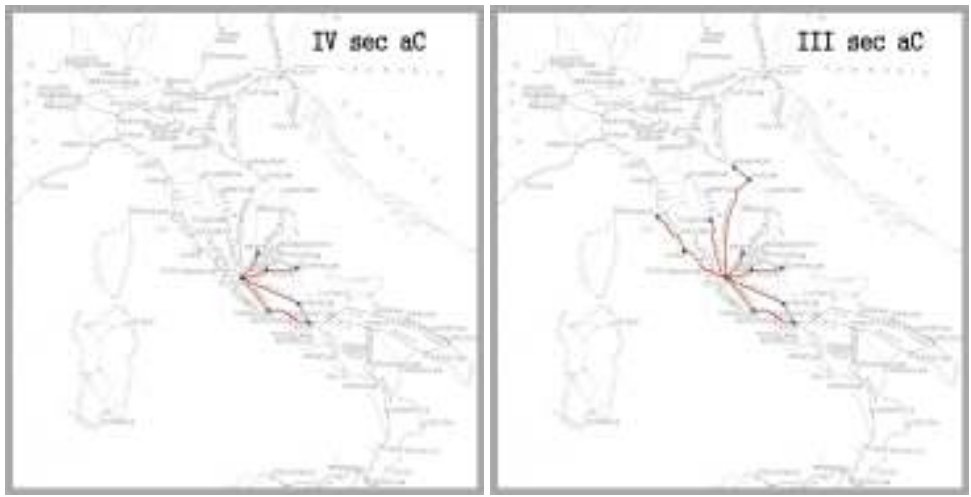

(a)        (b)

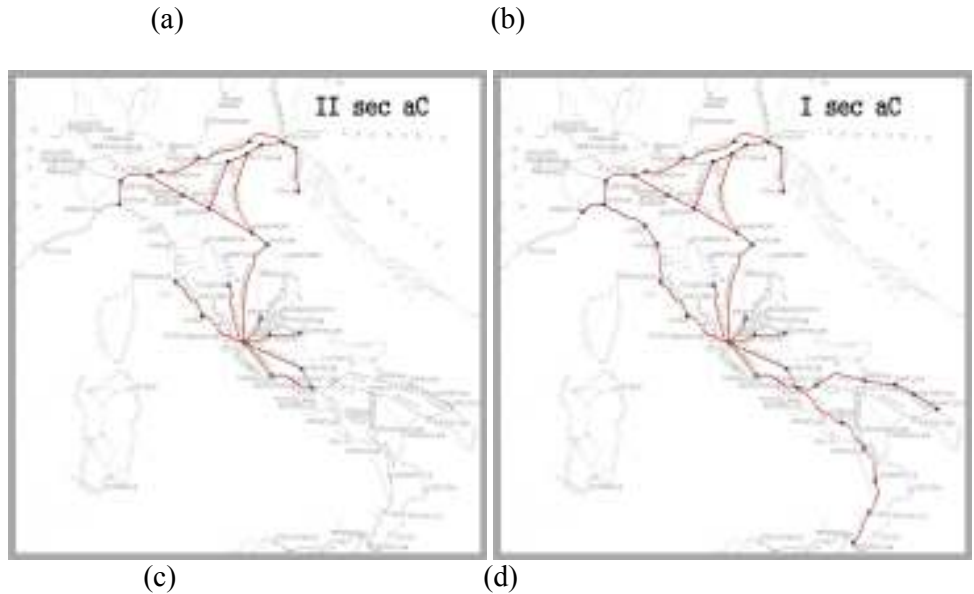

(c)        (d)

**FIGURE 10. Proposed stages of Roman street network development based on experiments with *P. polycephalum*. Routers of possible propagation are shown by red/dark-grey superimposed on the map of Iron Age Italy 1000-100 BC [35].**

# 5. Conclusions

Experimental laboratory studies and computer simulations confirmed that *P. polycephalum* match the street network to a certain degree. Based on the laboratory experiments we can propose the following sequence of road network development in Italian peninsula. At the initial stage Roma is linked with Corfinium, Capua and Reate (Fig. 10a). Next stage is characterised by northbound development: the roads are built to Populonia, Clusium and Ariminum (Fig. 10b). The propagation continuation towards northern parts till Roma is connected with Genua, Placentia, Verona, Aquileia and Pola (Fig. 10c). After the northern part of Italian peninsula is well serviced with roads the development starts southward. The roads are built from Capua to Brundisiu and Rhegnium (Fig. 10b). Such development is substantially matched by dynamics of a spanning tree growing from Roma 11.

By comparing the road network and slime mould networks with proximity graphs we found that in this particular of Italian peninsula the road networks and slime mould networks show high degree of affinity to spanning trees and relative neighbourhood graphs (see also [36,37] for use of proximity graphs in simulation of urban networks). This could be a sign that Roman transportation system, developed in ancient times, was based on strict mathematical logic but also possessed all features of bio-logic, at least as related to reasoning at the level of amorphous spatially distributed living creatures.

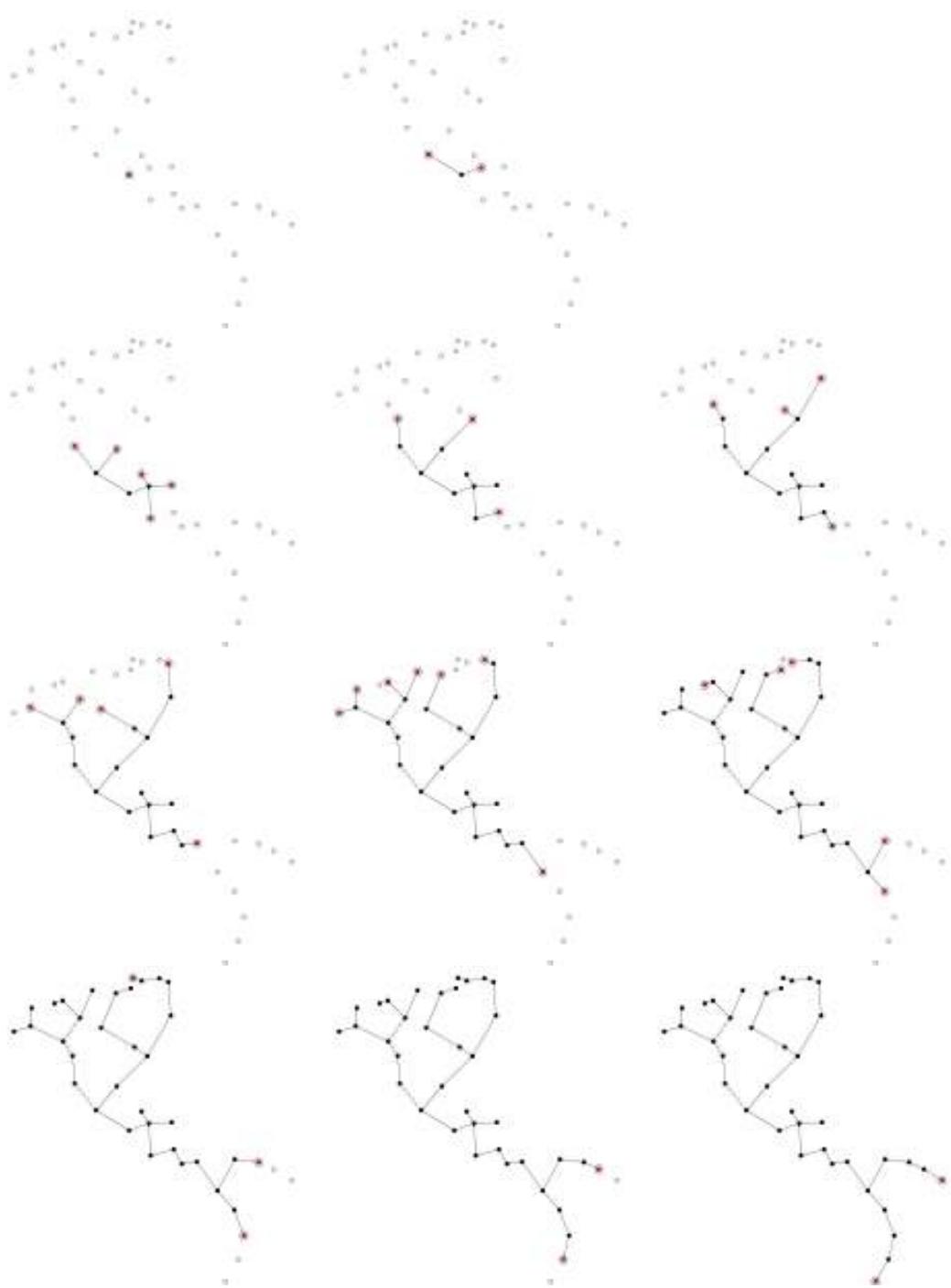

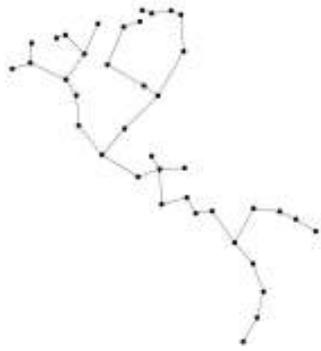

**Figure 12. Growth of a spanning tree routed in Roma, on cities used as nodes in illustration Fig. 10. Nodes occupied by "active zones", or growth cones of the tree, are encircled.**

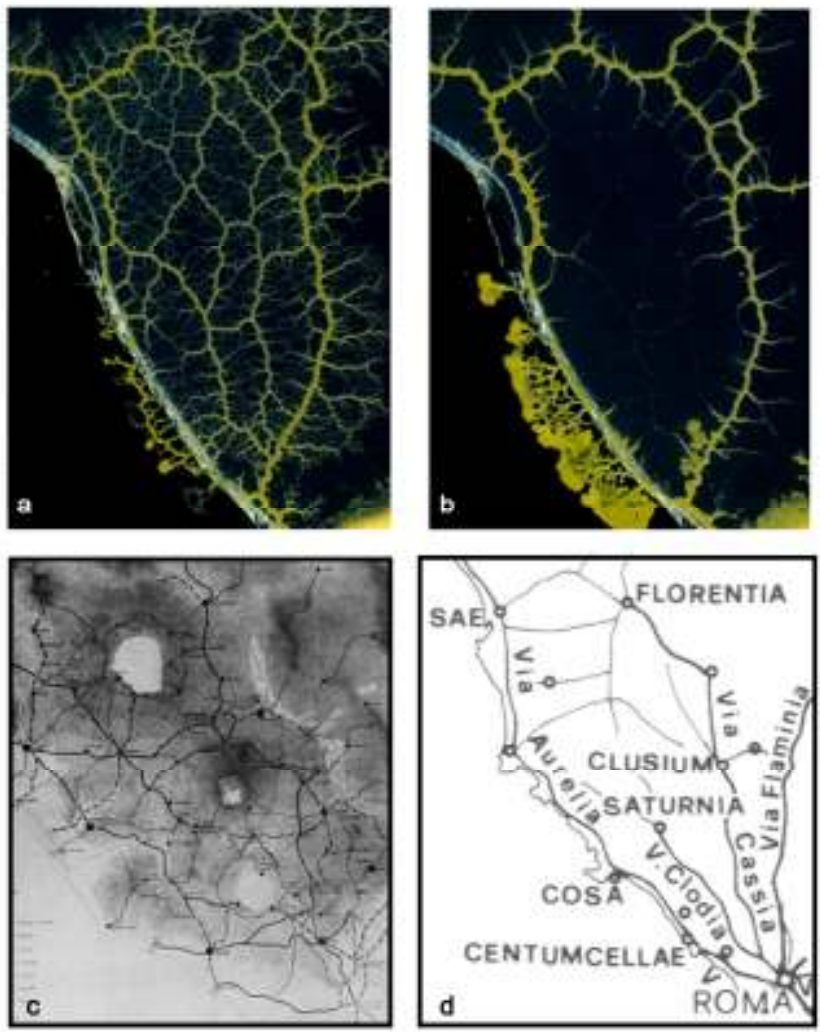

**FIGURE 12. Similarity between evolution of road networks in Italian peninsula and corresponding protoplasmic networks. (a) Early protoplasmic tube network, (b) After tube network adaptation (20m), (c) Primordial Etrurian trails (approx V Sec b.C.) [35], (d) Imperial Roman Street Networks (I Sec. a.C) [35].**

The results are encouraging and inspire us to undertake further investigations. For example, during laboratory experiments we noticed striking similarities between sprouting, scouting and space-searching executed by *P. polycephalum* and evolution of transport networks in the Tuscany region of the Italian peninsula (Fig. 12). The pre-Roman road networks in this region show signs of rather extensive exploration of the space (Fig. 12c). The same strategy is explored by slime mould during its initial stage of substrate colonisation (Fig. 12a). With time — due to implicit competition between roads — only a few roads are selected by the masses of travellers, as e.g. Etrurian paths in the Tuscany region (Fig. 12d). The same selection of few transport links occurs in protoplasmic networks of *P. polycephalum* (Fig. 12b) as protoplasmic tubes with lesser flux are eradicated. We believe our experiments will lead to further extensive experiments on bio-inspired urban developments boosted by paleotopography, particle-based modes of large-scale collectives and nature-inspired computing.